\documentclass[twocolumn]{aastex62} % The type of document

\usepackage{graphicx}

\usepackage{float}

\usepackage{natbib}

\usepackage{amsmath}
\usepackage{gensymb}
\usepackage{listings}
\usepackage{color}
\usepackage{amssymb}
\usepackage{multirow}
\usepackage{mathtools}
\usepackage{filecontents}
\usepackage{xcolor}
\usepackage{environ}
\usepackage{expl3}
\usepackage{bm}
\usepackage{textcase}
\usepackage{url}

\begin{document}
\title{Synchrotron maser from weakly magnetised neutron stars as the emission mechanism of fast radio bursts}
\author{Killian Long}
\author{Asaf Pe'er}
\affiliation{Department of Physics, University College Cork, Cork, Ireland; \url{killian.long@umail.ucc.ie} }

	\begin{abstract}
		The origin of Fast Radio Bursts (FRBs) is still mysterious. All FRBs to date show extremely high brightness temperatures, requiring a coherent emission mechanism. Using constraints derived from the physics of one of these mechanisms, the synchrotron  maser, as well as observations, we show that accretion induced explosions of neutron stars with surface magnetic fields of $B_*\lesssim10^{11}$ G are favoured as FRB progenitors.
	\end{abstract}
	

	\section{Introduction}
	\label{sec:intro}
	
	Fast Radio Bursts (FRBs) are bright radio transients of millisecond duration. A total of 33 FRBs have been published to date \citep{2016PASA...33...45P}\footnote{http://www.frbcat.org/}. They have typical fluxes of $\sim1\,\text{Jy}$ and are distinguished by their large dispersion measures (DM). These are in the range $176\,\text{pc cm}^{-3}$ to $2596\,\text{pc cm}^{-3}$, an order of magnitude greater than values expected from Milky Way electrons \citep{2016PASA...33...45P,2016MNRAS.460L..30C}, suggesting an extragalactic origin for FRBs.\par 
	 Of the 33 FRBs to date, 32 show no evidence of repetition. However, one of the bursts, FRB 121102, has been observed to repeat, allowing it to be localised \citep{2016Natur.531..202S}. A persistent counterpart and host galaxy were identified at a redshift of $z=0.193$, equivalent to a luminosity distance of $d_L=972\,$Mpc, strengthening the case for an extragalactic origin for FRBs \citep{2017ApJ...834L...7T,2017ApJ...834L...8M,2017Natur.541...58C}. \par 
	  The nature of FRB progenitors is still unknown. The small scale and large energies involved has led most models to consider compact objects such as neutron stars playing crucial roles in the production of FRBs. The numerous proposed models fall into two classes, cataclysmic and non-cataclysmic.  Cataclysmic models include 'blitzars' (collapsing neutron stars) \citep{2014AaA...562A.137F}, binary neutron star mergers \citep[e.g.][]{2012ApJ...755...80P}, white dwarf mergers \citep{2013ApJ...776L..39K} and neutron star-black hole mergers \citep{2015ApJ...814L..20M}. Non-cataclysmic models include giant pulses from extragalactic pulsars and young neutron stars \citep[e.g.][]{2012MNRAS.425L..71K, 2016MNRAS.457..232C} and flares from soft gamma repeaters (SGRs) \citep{2013arXiv1307.4924P,2014MNRAS.442L...9L,2017ApJ...843L..26B}. \par
	  Despite the large degree of uncertainty regarding the nature of the progenitor, there is a consensus on the need for a coherent emission process. This follows from the extremely high brightness temperatures of up to $T_b\sim 10^{37}\,\text{K}$ \citep{2016MPLA...3130013K}. As we show here, this requirement holds the key to understanding the nature of the progenitor. Analysing the conditions required to produce the necessary coherent emission allows us to place strong constraints on possible FRB progenitors. \par
	  We find that the conditions found in the environments of neutron stars with surface magnetic fields of $B_*\lesssim10^{11}$ G are similar to those required for a coherent emission mechanism, the synchrotron maser, to produce a FRB. Furthermore, the proportion of neutron stars with these magnetic fields is $\sim10\%$, and the FRB rate is comparable to this fraction of the neutron star formation rate. These results allow us to propose weakly magnetised neutron stars as FRB progenitors.

\section{The Basic Physics of the Synchrotron Maser}
\label{sec:maser}
 	Several mechanisms have been proposed to explain the coherent emission required to produce the extreme brightness temperatures of FRBs. Models include coherent curvature emission \citep{2017MNRAS.468.2726K,2017arXiv170807507G}, the cyclotron/synchrotron maser \citep{2014MNRAS.442L...9L,2017ApJ...843L..26B,2017ApJ...842...34W,2017MNRAS.465L..30G} and collisionless Bremsstrahlung in strong plasma turbulence \citep{2016PhRvD..93b3001R}.
 	
 	 Here we examine the synchrotron maser as the mechanism responsible for FRBs. The maser has the advantage of being a viable emission mechanism over a range of magnetic fields and number densities, as well as not requiring particles to be bunched in small volumes in order to obtain coherent emission \citep{2017MNRAS.465L..30G}. Previous works invoking the maser have examined specific models \citep{2014MNRAS.442L...9L,2017ApJ...843L..26B,2017ApJ...842...34W} or the mechanism itself \citep{2017MNRAS.465L..30G}, but have not used the mechanism's properties to derive general constraints on the progenitor.  \par
Maser emission is produced due to interaction between electromagnetic waves and energetic particles in a plasma, which can result in negative absorption and stimulated emission under certain conditions \citep{1985SSRv...41..215W,2006AaARv..13..229T}. The behaviour of the maser is determined by the form of the particle distribution and the environment where it occurs. For masing to occur, a population inversion in the electron distribution is required \citep{1985SSRv...41..215W,1979rpa..book.....R}.

 Maser emission has been suggested to occur in astrophysical sources for two different types of environments, differentiated by whether the plasma magnetisation is greater or less than unity, as different mechanisms are responsible in the two cases. The magnetisation can be quantified by the ratio $\nu_p/\nu_B$, where $\nu_p$ is the plasma frequency, given by  $\nu_p=\sqrt{ne^2/\pi\gamma m_e}$. Here $n$ is the number density of the plasma and $\gamma$ is the Lorentz factor of the electrons. The gyration frequency of the plasma particles, $\nu_B$, is given by $\nu_{B}=eB/(2\pi\gamma m_ec)$.

Here we examine both cases. In scenario (i) we investigate a homogeneous magnetised plasma ($\nu_p/\nu_B<1$), with a constant ambient magnetic field $B$. The plasma consists of a cold background component, which supports the propagation of the waves, and a less dense nonthermal component. The emission is due to gyroresonant interactions between the electrons and electromagnetic waves \citep{1985SSRv...41..215W}. In this scenario, we consider the nonthermal component to be a mildly relativistic magnetised plasma \citep{1986A&A...165..211L}, rather than the nonrelativistic magnetised plasma which has been proposed as the source of phenomena such as auroral kilometric radiation (AKR) in Earth's aurora, as well emission from other planets, the Sun, and blazars \citep{2006AaARv..13..229T, 2005ApJ...625...51B}. This mechanism is not applicable to highly relativistic plasmas, as masing can only occur when individual harmonics do not overlap \citep{1985PPCF...27.1037R, 1990PhFlB...2..867Y}. At higher Lorentz factors, the emission can be described by the synchrotron approximation \citep[e.g.][]{1982ApJ...259..350D}.

 In scenario (ii) we consider a weakly magnetised ($\nu_p/\nu_B>1$) relativistic nonthermal plasma. Maser emission in these conditions has been proposed as the source of radio emission from gamma ray burst afterglows \citep{2002ApJ...574..861S}. In this scenario the masing emission is due to the Razin effect, a modification of the emission from a relativistic plasma with respect to the vacuum case, which can result in either suppression or, when a population inversion is present, amplification of the emitted signal \citep{1966Sci...154.1320M,1967SvA....11...33Z}. This is due to a change in the beaming angle of the radiation when the refractive index of the plasma is less than unity \citep{1979rpa..book.....R}. For this case, a relativistic plasma is required, as the Razin effect is a relativistic effect and so would not effect cyclotron emission. This restriction does not apply in the magnetised case, due to the Razin effect only being relevant for $\nu_p/\nu_B>1$, as emission at the Razin frequency of $\nu_R*\approx\nu_p \text{min}\left\lbrace\gamma,\sqrt{\nu_p/\nu_B}\right\rbrace$ would otherwise not be visible.

\section{Physical and observational constraints of the allowed parameter space region that enables the production of FRBs}

\label{sec:constrain}
The physical conditions in the region where the masing takes place can be constrained using the physics of the maser and constraints from observations, allowing us to place limits on the magnetic fields and number densities where the synchrotron maser can plausibly be the emission mechanism for FRBs. 

 We consider a cataclysmic FRB progenitor. However, as the repeating burst FRB 121102 is the only one to have a known redshift, it is the only source which provides observational constraints for quantities such as the burst energy and the DM of the host galaxy. Therefore, we use the values it provides as representative limits for our calculations. 
 
 The data enable us to obtain constraints linking the size and number density of the masing region to the magnetic field of the neutron star. These constraints are obtained from: (i) the energetics of the burst and size of the masing region, (ii) the efficiency of the maser mechanism, (iii) the dispersion measure of the burst and (iv) the frequency of the signal.
 
  We consider that masing takes place in a spherical shell of thickness $d$, located a distance $R$ from the central object. The maser will be activated by the formation of a population inversion in the shell. The magnetic fields and short timescale ($\lesssim d/c\Gamma$) required suggest this object is a neutron star, though the timescale for maser emission is given by the duration of the maser itself \citep{2017MNRAS.465L..30G}. Assuming a relativistic blast wave the shocked plasma has a width $\sim R/\Gamma$ \citep{1976PhFl...19.1130B}, where $\Gamma$ is the Lorentz factor of the blast wave. This typical width provides the first constraint, on the thickness of the shell:

\begin{equation}
d\sim\dfrac{R}{\Gamma}\text{ .}
\label{eq:1}
\end{equation}

 The minimum thickness of the shell depends on the number of particles that contribute to the masing, $N_{e}=E/(\eta \left<E_e\right>)$, and their number density, $n_{e}$. Here, $E$ is the energy of the bursts (in the range $10^{38}-10^{40}\,\text{erg}$ for the repeater \citep{2017ApJ...834L...7T,2017ApJ...850...76L}), $\eta$ is the fraction of the electrons' energy that contributes to the maser and $\left<E_e\right>$ is the average energy of the masing electrons. These shocked electrons have a thermal energy of $\gamma\approx\Gamma$.
 
 Constraint (ii) originates from the efficiency of the maser. For the maser to be a viable emission mechanism, the growth rate of the signal must be large enough to extract a fraction $\eta$ of the particle energy. The maser will be quenched when the maser reaches saturation. The efficiency of the maser mechanism in simulations of relativistic shocks was shown to be $\eta\lesssim10^{-1}$ \citep[e.g.][]{1992ApJ...391...73G, 2009ApJ...698.1523S}. In the case of AKR, the efficiency is in the range $\eta\sim10^{-1}-10^{-3}$ \citep{1985SSRv...41..215W}. The exact value, however, depends on the form of the particle distribution, which is uncertain. We therefore examine values of $\eta$ in the range $10^{-3}\lesssim\eta\lesssim10^{-1}$ in this work. \citet{2017arXiv171010270L} give upper limits to the efficiency of $\eta\lesssim10^{-5}$, derived from limits on the brightness temperature from induced Compton scattering. However, plasma experiments suggest that this saturation effect is not observed for high $T_B$ \citep{2016PhRvD..93b3001R, 1998MNRAS.301..414B}.\par The growth rate, and therefore the efficiency, depends on the distribution function of the electrons. There is a wide range of possible distributions which can provide the requisite population inversion. We do not specify an exact form for the distribution as our results are unchanged provided the growth rate is large enough to extract the required energy over the width of the masing cavity.
 
 The third constraint comes from the dispersion measure. Assuming the DM from the source is solely due to the particles in the shell, one has $DM_{source}=DM_{shell}$, where $DM_{shell}=n_cd$ for a cold plasma and $DM_{shell}=n_ed/2\gamma$ for a relativistic plasma. Here, $n_c$ and $n_e$ denote the cold and relativistic electrons in the shell \citep{2008ApJ...688..695S}. The contribution to the DM from the source region is uncertain. The total DM value also contains contributions from the Milky Way, Milky Way halo, the intergalactic medium (IGM) and the host galaxy. For FRB 121102, \citet{2017ApJ...834L...7T} estimate the DM due to the host galaxy as $55\lesssim DM_{host}\lesssim 225\,\text{pc cm}^{-3}$. The contribution to this from the galaxy rather than the source region depends on the location of the FRB within the galaxy. Using these values as guidelines, the DM due to the shell is $DM_{shell}\lesssim 225\,\text{pc cm}^{-3}$.

  Constraint (iv) is derived from equating the masing frequency to the emission frequency of the bursts, which have observed frequencies of approximately $\nu_{obs}\sim1.4\,\text{GHz}$. Here, one discriminates between the two scenarios. For weakly magnetised plasma, the maser frequency is given by the Razin frequency, $\nu_R*$, where the growth rate is at a maximum. Equating the Razin frequency to the emission frequency of $\nu_{obs}/\Gamma$ and noting $\gamma=\Gamma$ gives the magnetic field in the masing region as 
 \begin{equation}
 B_M=1.32\times10^{-13}\sqrt{n_e^3\gamma^3}\text{ .}
 \label{eq:BM}
 \end{equation}
 
 \noindent The range
 
  \begin{equation}
  1<\frac{\nu_p}{\nu_B}<\gamma^2
 \label{eq:nrange}
 \end{equation} 
 
 \noindent delimits the range where $\nu_R*=\nu_p\sqrt{\nu_p/\nu_B}$. For $\nu_p/\nu_B>\gamma^2$ the Razin frequency is $\nu_R*=\gamma\nu_p$. While masing emission is still possible in this regime, the allowed parameter space is restricted to a small region with low magnetic fields due to constraints from the DM, shell size and observed emission frequency. Including the neutron stars in this region will not change the statistics for our model discussed below. The range of interest given by equation \ref{eq:nrange} can therefore be expressed in terms of the number density  using equation \ref{eq:BM} as
 
 \begin{equation}
 \dfrac{2.4\times10^{10}}{\gamma^3}<n_e<\dfrac{2.4\times10^{10}}{\gamma}\text{ ,} \label{eq:nupperlimit}
 \end{equation} 
 
 \noindent This provides upper and lower limits on the number density which depend on the Lorentz factor of the electrons.
 
On the other hand, for the strongly magnetised plasma the frequency of the maser is $\nu_M\approx l\nu_B$, giving a shell magnetic field of:

\begin{equation}
B_M\approx500l^{-1}\,G\text{ .}
\label{eq:freq}
\end{equation}

\noindent where $\nu_M=\nu_{obs}/\Gamma$ and $l$ is the harmonic number of the fastest growing mode.

\section{Results}
  For both the weakly and strongly magnetised plasmas, we investigate neutron star surface magnetic fields in the range $10^7\,\text{G}<B_*<10^{15}\,\text{G}$. This encompasses the full range of surface magnetic field values for all published pulsars \citep{2005AJ....129.1993M}\footnote{http://www.atnf.csiro.au/research/pulsar/psrcat}. In both cases the magnetic field outside the surface was taken to be of the form $B\propto1/r^3$ inside the light cylinder, and $B\propto1/r$ outside \citep{1969ApJ...157..869G}. The light cylinder radius $r_L=cP/2\pi$ is the radius at which the co-rotating speed is equal to the speed of light. Here $P$ is the period of the pulsar.  We also investigate the full range of number densities in the masing region. \par
\begin{figure}
  	\centering
  	\includegraphics[width=\columnwidth]{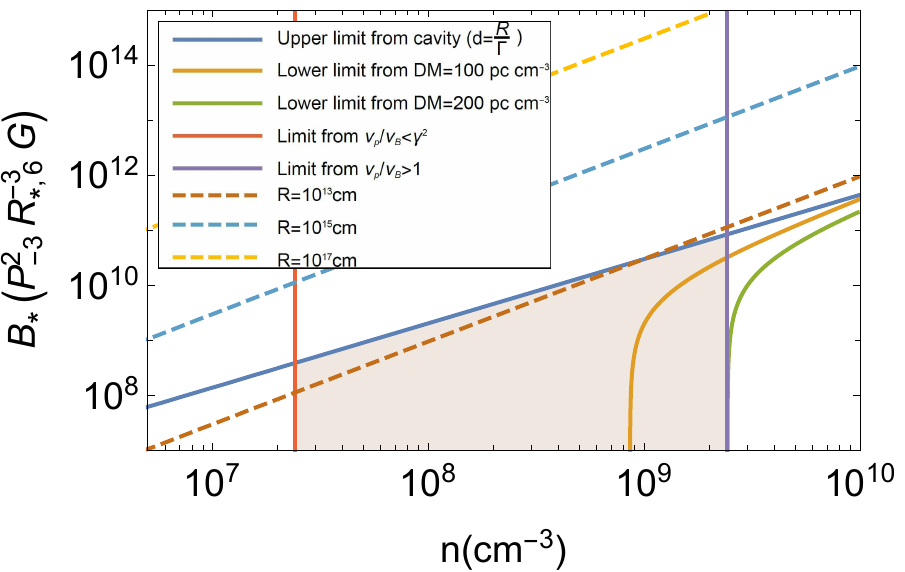}
  	\caption{Parameter space (shaded region) for the synchrotron maser with $\nu_p/\nu_B>1$, $\gamma=10$, $E=10^{40}$ erg and $\eta=10^{-3}$. Solid lines show limits while the dashed lines show lines of constant radius. Values of $B_*\lesssim10^{10}\,\text{G}$, $n\sim10^8\,\text{cm}^{-3}$ and $R\sim10^{13}\,\text{cm}$ are preferred. For larger DM values the lower limit will decrease. Increasing the value of $\gamma$ results in less restrictive DM constraints, lower allowed number densities and higher allowed neutron star surface magnetic field values.}
  	\label{fig:2}
  \end{figure}

 \textbf{In the weakly magnetised scenario}, the range of $n_e$ is given by equation \ref{eq:nupperlimit}. Lorentz factors of $\gamma=2,5,10,100,10^{3}\,\text{and}\,10^6$ were examined. The larger values were chosen to examine conditions similar to pulsar wind nebulae, which have Lorentz factors of up to $\sim10^6$ \citep{2006ARAaA..44...17G, 2009ASSL..357..421K}. The allowed parameter space for $\gamma=10$, $E=10^{40}$ erg and $\eta=10^{-3}$ is shown in Figure \ref{fig:2} as an example. In this case, the results indicate that the allowed parameter space is restricted to low magnetic fields and $n\sim10^8\,\text{cm}^{-3}$. Increasing the allowed values of the DM results in the lower limit decreasing. 
  
  The allowed surface magnetic field values depend on the Lorentz factor, number density in the masing region and the distance to the masing region, $R$. As the volume of the shell is $V\approx4\pi R^2d$, the distance to the masing region and the number density are related by the expression \begin{equation}
  R\approx\left(\frac{E}{4\pi\eta mc^2n_e}\right)^{1/3}\,\text{.}
  \end{equation}
  
  \noindent Using equation \ref{eq:BM}, the surface magnetic field can therefore be expressed as $B_*\propto n_e^{7/6}\gamma^{3/2}$.
  Taking into account the maximum allowed number density from equation \ref{eq:nupperlimit}, $n_{e,max}\propto \gamma^{-1}$, the maximum surface magnetic field is $B_{*,max}\propto\gamma^{1/3}$. Thus, the allowed surface magnetic field depends only weakly on the Lorentz factor. Therefore, even for very large values of $\gamma$ only low values of $B_*$ are attainable.

    We find that a neutron star with a surface magnetic field of $B_*\lesssim10^{10}-10^{11}\,\text{G}$ is required for emission at the appropriate frequency and energy, increasing to $B_*\lesssim10^{12}\,\text{G}$ only in the extremely relativistic $\gamma=10^6$ case. Ruling out pulsars with magnetic fields greater than $10^{11}$ G leaves approximately $14.5\%$ of the total population \citep{2005AJ....129.1993M}.  For FRBs with lower energy and greater efficiency, the upper limit on the magnetic field can be significantly lower at $\sim10^{9.5}$ G. Less than $10\%$ of pulsars have magnetic fields lower than this value. These upper limits on $B_*$ are thus very strong constraints as they rule out the majority of pulsars as being possible hosts for the synchrotron maser in the context of FRBs. The $\sim15\%$ of the known pulsar population that meet the criteria are therefore candidates to be FRB progenitors. Therefore, the FRB rate should be a similar fraction of the neutron star formation rate. The neutron star formation rate is approximated by the core-collapse supernova rate which is approximately $\mathcal{R}_{SN}\sim(1.42\pm0.3)\times10^5\,\text{Gpc}^{-3}\,\text{yr}^{-1}$ \citep{2009AaA...499..653B}, while the rate of FRBs is approximately $\mathcal{R}_{FRB}\sim0.98^{+1.15}_{-0.89}\times10^4\,\text{Gpc}^{-3}\,\text{yr}^{-1}$ \citep{2016MNRAS.460L..30C}. The ratio of the two rates is $\mathcal{R}_{FRB}/\mathcal{R}_{SN}\sim0.07$.
  This value is similar to the fraction of pulsars with surface magnetic fields of less than $10^{10}\,\text{G}$, which is $\sim0.1$. 
  
  The limits obtained from $DM_{shell}$ also constrain our results significantly. They have a particularly marked effect in the cases with larger numbers of particles in the masing region. For the higher energy bursts the DM limits severely constrain the cases with lower Lorentz factors, while for the lower energy bursts they are only relevant for $\Gamma=2$. The lower limit on $n_e$ depends on the case under consideration. Bursts with higher energies and Lorentz factors have lower allowed number densities. The lowest density of $\sim1\,\text{cm}^{-3}$ was achieved for $\gamma=10^6$.
  
 \textbf{For the highly magnetised plasma scenario}, we examine background (cold electron) number densities of up to $n_c=10^7\,\text{cm}^{-3}$, as values larger than this were ruled out by constraints from the DM. For each value of $n_c$, we examine the range $10^{-3}<\frac{n_e}{n_c}<10^{-1}$, where the lower limit is set by the luminosity requirements. The growth rate decreases with $n_e$, and so smaller values result in growth rates which are too low to produce the required luminosity. A Lorentz factors of $\gamma=2$ was examined, as the maser in this case is not relevant in highly relativistic scenarios. At the maximum number density of $n_c\sim10^{7}\,\text{cm}^{-3}$, the upper limit on the surface magnetic field is $\sim10^{12}\,\text{G}$. As $B_*\propto n^{-1/3}$, at low number densities higher magnetic fields are obtainable. However, this scenario can be ruled out entirely through constraints obtained from the physics of the blast wave.

 The Lorentz factor of a blast wave expanding into the interstellar medium (ISM) is given by \begin{equation}
 \Gamma=\left(\frac{17E}{16\pi n_{ISM}m_pc^2R^3}\right)^{1/2}\,\text{,}
 \end{equation}
 \noindent where $E$ is the energy of the blast wave, $m_p$ is the proton mass, $n_{ISM}$ is the density of the ISM and $n_e<<n_c$ \citep{1976PhFl...19.1130B}. This gives the distance to the shell as
 $R_{15}\lesssim1.31 E_{40}^{1/3}n_{ISM,0}^{-1/3}\eta_{-3}^{-1/3}\Gamma_0^{-2/3}\,\text{cm}$, where $Q=10^xQ_x$ in cgs units. Using equation \ref{eq:freq} and $B_*\approx\frac{c^2P^2B_MR}{4\pi^2R_*^3}$, this condition restricts the surface magnetic field to
 \begin{equation}
 B_{*,13}\lesssim1.49E_{40}^{1/3}n_{ISM,0}^{-1/3}\eta_{-3}^{-1/3}\Gamma_0^{-2/3}R_{*,6}^{-3}P_{-3}^2l^{-1}\,\text{G .}
 \label{eq:b1}
 \end{equation}  
 \noindent Here, $R_*$ is the radius of the neutron star.  However, the number density is also related to $R$ and $B_*$ through equation \ref{eq:1}, resulting in the condition
 
 \begin{equation}
  B_{*,13}\approx11.3 E_{40}^{1/3}n_{ISM,0}^{-1/3}\eta_{-3}^{-1/3}R_{*,6}^{-3}P_{-3}^2l^{-1}\,\text{G.}
  \label{eq:b2} 
 \end{equation}
 
 \noindent Equation \ref{eq:b2} does not satisfy the condition in equation \ref{eq:b1} for any value of $\Gamma$. Therefore the maser in a strongly magnetised plasma can be ruled out as the possible emission mechanism.

  \section{Discussion}
    Emission from the synchrotron maser can be circularly, elliptically or approximately linearly polarised, depending on the electron distribution function and plasma parameters \citep{2006AaARv..13..229T,2002ApJ...574..861S}. Similarly, both circular \citep[e.g. ][]{2015Natur.528..523M} and linear polarisation \citep[e.g. ][]{2018Natur.553..182M} has been measured in FRB observations. However, the heterogeneous nature of FRB polarisation measurements to date makes it difficult to draw useful constraints from the data. \par
   The density constraints obtained from the maser can be compared to the densities found in the vicinity of neutron stars. In the case of pulsar wind nebulae, densities of $n\sim10^{-6}\,\text{cm}^{-3}$ and magnetic fields of $B_M\sim10^{-2}-10^{-1}\,\text{G}$ are expected \citep{2014MNRAS.442L...9L,2009ASSL..357..421K,2006ARAaA..44...17G,2014MNRAS.438.1518O}. Neither of these values lie within the allowed parameter space for the synchrotron maser, ruling out this scenario.
   
    In order to account for the larger density values required by our constraints, we are lead to suggest a scenario where weakly magnetised neutron stars undergo an accretion induced explosion \citep{1994ApJ...437..727K}. The material expelled by this explosion can then form a shell of width $\sim R/\Gamma$ in which a population inversion is formed, and as a result masing takes place. Accreting neutron stars in low mass X-Ray binaries (LMXBs) have typical wind densities of $10^{13}-10^{15}\,\text{cm}^{-3}$ at radii of approximately $10^{10}\,\text{cm}$ \citep{2016AN....337..368D}. While these density values are too high for the maser, our scenario considers the masing emission to occur at larger distances of $R\sim10^{13}\,\text{cm}$. As, at constant velocity, $n\propto r^{-2}$, the particles from the accretion induced explosion could plausibly provide suitable number densities for the maser at these distances. Pulsars in binary systems with $B_*<10^{11}\,\text{G}$ have typical periods of $\sim \text{few ms}$ and make up $\sim0.09$ of the total population \citep{2005AJ....129.1993M}, comparable to the ratio of the FRB and neutron star formation rates. As a result, this scenario would require a significant fraction of low magnetic field neutron stars in binaries to undergo such an event due to the similarities between the FRB rate and the neutron star formation rate. The scenario where the masing occurs in a strongly magnetised plasma is ruled out due to the impossibility of obtaining a blast wave of sufficient velocity at the required radius and number density.

\acknowledgments
KL acknowledges the support of the Irish Research Council through grant number GOIPG/2017/1146. The authors also thank the referee for their useful comments.

\end{document}